\documentclass[aps,prd,onecolumn,showpacs,showkeys,amsmath,amssymb]{revtex4}
\usepackage{epsfig} 
\usepackage{amsmath} 
\usepackage{makecell}
\usepackage{array}
\usepackage{subfigure}
\usepackage{bm}
\usepackage{longtable}
\usepackage{tabularx}
\usepackage{booktabs}
\usepackage{appendix}
\usepackage{caption2}
\usepackage{amsfonts}
\usepackage{amsmath}
\usepackage{graphicx}
\graphicspath{{figs/}}
\usepackage{subfigure}
\usepackage{dcolumn}
\usepackage{siunitx}
\usepackage{bm}
\usepackage{booktabs}
\usepackage[utf8]{inputenc}
\usepackage{float}
\usepackage{longtable,lscape}
\usepackage{txfonts}
\usepackage{overpic}
\usepackage{amssymb}
\usepackage{indentfirst}
\usepackage{epsfig}
\usepackage{feynmf}   %{feynmp}
\usepackage{epstopdf}   %{feynmp}
\usepackage{slashed}  %for Feynman symbols
\usepackage{multirow}
\usepackage{natbib}

\usepackage[colorlinks, citecolor=blue,anchorcolor=red,menucolor=red, linkcolor=red,filecolor=red,runcolor=red,urlcolor=blue,frenchlinks=red]{hyperref}

\usepackage{ulem}
\usepackage{color}

\makeatletter
\newcommand{\figcaption}{\def\@captype{figure}\caption}
\newcommand{\tabcaption}{\def\@captype{table}\caption}

\newcommand{\Rmnum}[1]{\expandafter\@slowromancap\romannumeral #1@}

\def\hlinewd#1{%
  \noalign{\ifnum0=`}\fi\hrule \@height #1 \futurelet
   \reserved@a\@xhline}
\makeatother

\newcommand\dqq{\Big\langle \bar q q \Big\rangle}
\newcommand\dss{\Big\langle \bar s s \Big\rangle}
\newcommand\dqGq{\Big\langle g_s \bar q \sigma G q \Big\rangle}
\newcommand\dsGs{\Big\langle g_s \bar s \sigma G s \Big\rangle}
\newcommand\dGG{\Big\langle \alpha_{s} GG \Big\rangle}

\begin{document}
\title{Light tetraquark states with exotic quantum numbers $J^{PC}=2^{+-}$}

\author{Qi-Nan Wang$^1$}
%\email{wangqn8@mail.sysu.edu.cn}
\author{Ding-Kun Lian$^1$}

\author{Wei Chen$^{1,\, 2}$}
\email{chenwei29@mail.sysu.edu.cn}
\affiliation{$^1$School of Physics, Sun Yat-Sen University, Guangzhou 510275, China\\ 
$^2$Southern Center for Nuclear-Science Theory (SCNT), Institute of Modern Physics, 
Chinese Academy of Sciences, Huizhou 516000, Guangdong Province, China}

\begin{abstract}
We study the masses of light tetraquark states $ud\bar{u}\bar{d}$ , $us\bar{u}\bar{s}$ and $ss\bar{s}\bar{s}$ with exotic quantum numbers $J^{PC}=2^{+-}$ using the method of QCD sum rules. It is found that there is no tetraquark operator with two Lorentz indices coupling to the $2^{+-}$ quantum numbers. To investigate such tetraquark states, we construct the interpolating tetraquark currents with three Lorentz indices and without derivative operator. We calculate the correlation functions up to dimension 10 condensates, and extract the $2^{+-}$ invariant functions via the projector operator. Our results show that the masses of the $ud\bar{u}\bar{d}$, $us\bar{u}\bar{s}$ and $ss\bar{s}\bar{s}$ tetraquark states with $J^{PC}=2^{+-}$ are about $3.3-3.5 ~\mathrm{GeV}$, $3.5-3.7 ~\mathrm{GeV}$ and  $3.67 ~\mathrm{GeV}$, respectively. We further discuss the strong decays of these light tetraquarks into the two-meson and baryon-antibaryon final states, and suggest to search for them in the $\rho\pi, \omega\pi, \phi\pi$, $b_{1}\pi$, $h_{1}\pi$, $K\bar K^\ast, K\bar{K}_{1}$, $\Delta\bar{\Delta}$, $\Sigma^{\ast} \bar{\Sigma }^{\ast}$, $\Xi^{\ast} \bar{\Xi }^{\ast}$, $\Omega\bar{\Omega }$ channels in the future.
\end{abstract}

%\date{\today}

\maketitle

\section{Introduction}
Quantum chromodynamics (QCD) is the fundamental theory to describe the strong interaction among quarks and gluons in a SU(3) gauge symmetry. In the traditional quark model, hadrons are $q\bar q$ mesons and $qqq$ baryons~\cite{GellMann:1964nj,Zweig:1964jf,ParticleDataGroup:2022pth}. However, QCD allows for the existence of hadron states beyond the quark model, such as tetraquarks, pentaquarks, hexaquarks, dibaryons, hybrid mesons, glueballs and so on~\cite{Meyer:2015eta,Chen:2016qju,Clement:2016vnl,Guo:2017jvc,Liu:2019zoy,Brambilla:2019esw,Chen:2022asf,Liu:2024uxn}. 

A compact tetraquark is made of a pair of diquark and anti-diquark. 
In QCD the light scalar mesons $\sigma(600)$, $\kappa(800)$, $a_{0}(980)$ and $f_{0}(980)$ have been considered as the candidates of light tetraquark states~\cite{Jaffe:1976ig,Black:1998wt,Maiani:2004uc,Chen:2007xr,Prelovsek:2008rf}. In 2006, the BaBar Collaboration observed $\phi(2170)$ in the $e^{-}e^{+}\rightarrow \phi f_{0}(980)$ process~\cite{BaBar:2006gsq}, which was confirmed by latter experiments~\cite{BES:2007sqy,BESIII:2014ybv,BESIII:2017qkh,BaBar:2007ptr,BaBar:2007ceh,BaBar:2011btv,Belle:2008kuo}. Since its observation, this vector resonance with $J^{PC}=1^{--}$ was considered as the candidate of a fully strange $ss\bar{s}\bar{s}$ tetraquark state~\cite{Deng:2010zzd,Wang:2006ri,Ke:2018evd,Chen:2018kuu}, although some other interpretations have not been excluded~\cite{Ho:2019org,Deng:2013aca}. Similarly, the $X(2239)$ structure observed by BESIII~\cite{BESIII:2018ldc} was also interpreted the light tetraquark state~\cite{Lu:2019ira,Azizi:2019ecm}.

The conventional $\bar{q}q$ mesons are forbidden to carry exotic quantum numbers, such as $J^{PC}=0^{--}$, $even^{+-}$, $odd^{-+}$. However, they can be reached in the tetraquark and hybrid meson configurations. To date, the only observed exotic $J^{PC}$ quantum numbers appear for the isovector $\pi_{1}(1400)$~\cite{IHEP-Brussels-LosAlamos-AnnecyLAPP:1988iqi}, $\pi_{1}(1600)$~\cite{E852:2001ikk} and $\pi_{1}(2015)$~\cite{E852:2004gpn} with $I^GJ^{PC}=1^-1^{-+}$ and the isoscalar $\eta_{1}(1855)$ with $I^GJ^{PC}=0^+1^{-+}$~\cite{BESIII:2022riz,BESIII:2022iwi}, in which $\pi_{1}(1400)$ and $\pi_{1}(1600)$ were also considered to be the same state~\cite{JPAC:2018zyd,COMPASS:2014vkj}. During the past several decades, these exotic  structures have been extensively investigated as the best candidates for hybrid mesons~\cite{Meyer:2015eta,Chen:2022asf,Chen:2022qpd,Chen:2023ukh,Chen:2022isv,Qiu:2022ktc,Shastry:2022upd,Shastry:2023ths,Shastry:2022mhk}. Nevertheless, they can also be interpreted as the light compact tetraquarks and hadronic molecules~\cite{Chen:2008qw,Chen:2008ne,Zhang:2019ykd,Huang:2022tpq,Yang:2022rck,Dong:2022cuw,Wan:2022xkx,Wang:2022sib,Su:2022eun,Yu:2022lwl,Yan:2023vbh}.
The light tetraquark states have also been studied for the $J^{PC}=0^{--}$~\cite{Jiao:2009ra,Huang:2016rro} and $0^{+-}$~\cite{Du:2012pn,Fu:2018ngx,Ray:2022fcl,Xi:2023byo} exotic channels. 

Recently, there are some theoretical investigations on the exotic hadrons with  $J^{PC}=2^{+-}$. In Ref.~\cite{Xi:2023byo}, the fully strange $ss\bar s\bar s$ tetraquark state with such quantum numbers was studied in QCD sum rules and its mass was calculated to be around 3.1 GeV. The nonstrange and strangeonium light one-gluon hybrid mesons with $J^{PC}=2^{+-}$ were studied by Lattice QCD and QCD sum rules to give the mass predictions about $2.4-2.7~\mathrm{GeV}$~\cite{Dudek:2011bn,Dudek:2013yja,Wang:2023whb}, which are much heavier than the results in the flux tube model~\cite{Isgur:1985vy}. In Refs.~\cite{Morningstar:1999rf,Chen:2005mg,Tang:2015twt,Chen:2021cjr}, the exotic $2^{+-}$ glueballs were investigated by Lattice QCD and QCD sum rules with diverse mass predictions. A new type of double-gluon hybrid mesons with exotic quantum numbers was proposed recently in QCD sum rules in Refs.~\cite{Chen:2021smz,Tang:2021zti,Su:2022fqr,Su:2023jxb,Su:2023aif,Lian:2024fsb}. 
In this work, we shall further study the mass spectra of the $ud\bar{u}\bar{d}$, $us\bar{u}\bar{s}$ and $ss\bar{s}\bar{s}$ tetraquark states with exotic quantum numbers $J^{PC}=2^{+-}$ by constructing the interpolating tetraquark currents with three Lorentz indices in the method of QCD sum rules.

This paper is organized as follows. In Sec.~\ref{Sec:2}, we construct the three-Lorentz-indices interpolating tetraquark currents that can couple to physical hadron states with $J^{PC}=2^{+-}$, and compose the projector operator to extract the invariant functions. In Sec.~\ref{Sec:3}, we calculate the correlation functions and spectral densities for these interpolating currents, and establish the mass sum rules for the $ud\bar{u}\bar{d}$, $us\bar{u}\bar{s}$ and $ss\bar{s}\bar{s}$ tetraquark systems. In Sec.~\ref{Sec:5}, the numerical analyses will be performed to obtain the tetraquark mass spectra. The last section is a brief summary and discussion.

\section{Interpolating Currents and Projectors}\label{Sec:2}
In this section, we construct the diquark-antidiquark interpolating currents coupling to the light tetraquark states with $J^{PC}=2^{+-}$. As a matter of fact, one can not construct a $J^{PC}=2^{+-}$ current with two Lorentz indices by using the Dirac gamma matrices only. To construct a current without derivative operator in such channel, one needs to consider the interpolating current with more than two Lorentz indices~\cite{Wang:2023whb}. We consider the diquark fields $q_a^T C \gamma_{5}q_b, q_a^T C q_b, q_a^T C \gamma_\mu \gamma_{5}q_b, q_a^T C \gamma_\mu q_b, u_a^T C \sigma_{\mu \nu} q_b, q_a^T C \sigma_{\mu \nu}\gamma_5 q_b$ and the corresponding antidiquark fields, where $a, b$ are color indices, and $T$ the transpose of the matrices. 

\begin{table*}[t]
  \centering
    \caption{The spins and parities of the diquark fields.}
    \renewcommand\arraystretch{1.6}
  \setlength{\tabcolsep}{1.em}{
  \begin{tabular}{ccc}			\hline \hline
  $q^{T}C\Gamma q$   & $J^{P}$   & States   \\    \hline
   $q_{a}^{T}C\gamma_{5} q_{b}$    & $ 0^{+}$    & $^{1}S_{0}$    \\
   $q_{a}^{T}C q_{b}$      & $ 0^{-}$        & $^{3}P_{0}$     \\
    $q_{a}^{T}C\gamma_{\mu}q_{b}$  &  $ 1^{+}$&    $ ^{3}S_{1} $  \\
     $q_{a}^{T}C\gamma_{\mu}\gamma_{5}q_{b}$    & $\Big\{\begin{array}{rl}
                0^{+} & \mbox{if }\mu=0  \\
               1^{-} & \mbox{if }\mu=1,2,3
            \end{array} $      &  $\begin{array}{rl}
                ^{1}S_{0}  \\
                ^{3}P_{1}
            \end{array} $  \\
      $q_{a}^{T}C\sigma_{\mu\nu}q_{b}$    & $\Big\{\begin{array}{rl}
                1^{-} & \mbox{if }\mu,\nu=1,2,3  \\
               1^{+} & \mbox{if }\mu=0,\nu=1,2,3
            \end{array} $      &  $\begin{array}{rl}
                ^{1}P_{1}  \\
                ^{3}S_{1}
            \end{array} $  \\
   $q_{a}^{T}C\sigma_{\mu\nu}\gamma_{5}q_{b}$    & $\Big\{\begin{array}{rl}
                1^{+} & \mbox{if }  \mu,\nu=1,2,3\\
               1^{-} & \mbox{if }\mu=0,\nu=1,2,3 
            \end{array} $      &  $\begin{array}{rl} ^{3}S_{1}  \\
               ^{1}P_{1}
            \end{array} $    \\
            \hline \hline
  \label{Tab:diquark}
  \end{tabular}}
\end{table*}

The spin-parities of the diquark fields with various Lorentz structures are shown in Table ~\ref{Tab:diquark}.
One can construct a tetraquark operator as
\begin{equation}
O_{ij}= (q_a^{T}C\Gamma_{i} q_b)(\bar{q}_{a}\Gamma_{j} C \bar{q}_{b}^{T})\, ,
\end{equation}
where the color structures of the diquark and antidiquark fields depend on their Lorentz structures. It is easy to find the following identity under the charge conjugation transform~\cite{Chen:2010ze}
\begin{equation}
  \mathbb{C} O_{ij} \mathbb{C}^{-1}=O_{ij}^T\, .
\end{equation}
One can find the tetraquark operators with even and odd $C$-parities as 
\begin{equation}
    S=O_{ij}+O_{ij}^T\,, \quad A=O_{ij}-O_{ij}^T\,.
\end{equation}
Considering a tetraquark operator with two symmetric Lorentz indices $O_{ij}=O_{\{\mu,\nu\}}$, it can couple to the spin-2 hadron state with the following  possible Lorentz structures without derivative 
\begin{equation}
  \{\Gamma_{i},\Gamma_{j}\}=\{\gamma_{\mu},\gamma_{\nu}\},\{\gamma_{\mu}\gamma_{5},\gamma_{\nu}\gamma_{5}\},\{\sigma_{\mu\alpha},\sigma_{\alpha\nu}\},\{\sigma_{\mu\alpha}\gamma_{5},\sigma_{\alpha\nu}\gamma_{5}\},\{\gamma_{\mu}\gamma_{5},\gamma_{\nu}\},\{\sigma_{\mu\alpha}\gamma_{5},\sigma_{\alpha\nu}\}\,.
\end{equation}
However, all of these tetraquark operators can not achieve the exotic quantum numbers $J^{PC}=2^{+-}$. 
In this work, we construct the following $2^{+-}$ interpolating tetraquark currents with three Lorentz indices 
\begin{equation}
  \begin{aligned}
    & J_{\alpha \mu \nu}^{1}=u_a^T C\gamma_{\alpha}  d_b\left(\bar{u}_a \sigma_{\mu \nu} C \bar{d}_b^T-\bar{u}_b \sigma_{\mu \nu} C \bar{d}_a^T\right) - u_a^T C \sigma_{\mu \nu} d_b\left(\bar{u}_a\gamma_{\alpha} C \bar{d}_b^T-\bar{u}_b\gamma_{\alpha} C \bar{d}_a^T\right), \\
    & J_{\alpha \mu \nu}^{1\prime}=u_a^T C\gamma_{\alpha} d_b\left(\bar{u}_a \sigma_{\mu \nu} C \bar{d}_b^T+\bar{u}_b \sigma_{\mu \nu} C \bar{d}_a^T\right) - u_a^T C \sigma_{\mu \nu} d_b\left(\bar{u}_a\gamma_{\alpha} C \bar{d}_b^T+\bar{u}_b\gamma_{\alpha} C \bar{d}_a^T\right), \\
    & J_{\alpha \mu \nu}^{2}=u_a^T C\gamma_{\alpha} \gamma_{5}d_b\left(\bar{u}_a \sigma_{\mu \nu} C \bar{d}_b^T-\bar{u}_b \sigma_{\mu \nu} C \bar{d}_a^T\right) - u_a^T C \sigma_{\mu \nu} d_b\left(\bar{u}_a\gamma_{\alpha} \gamma_{5}C \bar{d}_b^T-\bar{u}_b\gamma_{\alpha} \gamma_{5}C \bar{d}_a^T\right), \\
    & J_{\alpha \mu \nu}^{2\prime}=u_a^T C\gamma_{\alpha} \gamma_{5}d_b\left(\bar{u}_a \sigma_{\mu \nu} C \bar{d}_b^T+\bar{u}_b \sigma_{\mu \nu} C \bar{d}_a^T\right) - u_a^T C \sigma_{\mu \nu} d_b\left(\bar{u}_a\gamma_{\alpha} \gamma_{5}C \bar{d}_b^T+\bar{u}_b\gamma_{\alpha} \gamma_{5}C \bar{d}_a^T\right),\\
    & J_{\alpha \mu \nu}^{3}=u_a^T C\gamma_{\alpha} d_b\left(\bar{u}_a \sigma_{\mu \nu} \gamma_{5}C \bar{d}_b^T-\bar{u}_b \sigma_{\mu \nu} \gamma_{5}C \bar{d}_a^T\right) - u_a^T C \sigma_{\mu \nu} \gamma_{5}d_b\left(\bar{u}_a\gamma_{\alpha} C \bar{d}_b^T-\bar{u}_b\gamma_{\alpha} C \bar{d}_a^T\right), \\
    & J_{\alpha \mu \nu}^{3\prime}=u_a^T C\gamma_{\alpha} d_b\left(\bar{u}_a \sigma_{\mu \nu} \gamma_{5}C \bar{d}_b^T+\bar{u}_b \sigma_{\mu \nu} \gamma_{5}C \bar{d}_a^T\right) - u_a^T C \sigma_{\mu \nu} \gamma_{5}d_b\left(\bar{u}_a\gamma_{\alpha} C \bar{d}_b^T+\bar{u}_b\gamma_{\alpha} C \bar{d}_a^T\right), \\
    & J_{\alpha \mu \nu}^{4}=u_a^T C\gamma_{\alpha} \gamma_{5}d_b\left(\bar{u}_a \sigma_{\mu \nu} \gamma_{5}C \bar{d}_b^T-\bar{u}_b \sigma_{\mu \nu} \gamma_{5}C \bar{d}_a^T\right) - u_a^T C \sigma_{\mu \nu} \gamma_{5}d_b\left(\bar{u}_a\gamma_{\alpha} \gamma_{5}C \bar{d}_b^T-\bar{u}_b\gamma_{\alpha} \gamma_{5}C \bar{d}_a^T\right), \\
    & J_{\alpha \mu \nu}^{4\prime}=u_a^T C\gamma_{\alpha} \gamma_{5}d_b\left(\bar{u}_a \sigma_{\mu \nu} \gamma_{5}C \bar{d}_b^T+\bar{u}_b \sigma_{\mu \nu} \gamma_{5}C \bar{d}_a^T\right) - u_a^T C \sigma_{\mu \nu} \gamma_{5}d_b\left(\bar{u}_a\gamma_{\alpha} \gamma_{5}C \bar{d}_b^T+\bar{u}_b\gamma_{\alpha} \gamma_{5}C \bar{d}_a^T\right)\, ,
   \end{aligned}  
\label{Eq:current}
\end{equation}
in which the currents $J_{\alpha \mu \nu}^{1}$, $J_{\alpha \mu \nu}^{2}$, $J_{\alpha \mu \nu}^{3}$ and $J_{\alpha \mu \nu}^{4}$ have the color structure $\mathbf{3} \otimes \bar{\mathbf{3}}$, while $J_{\alpha \mu \nu}^{1\prime}$, $J_{\alpha \mu \nu}^{2\prime}$, $J_{\alpha \mu \nu}^{3\prime}$ and $J_{\alpha \mu \nu}^{4\prime}$ have the color structure $\bar{\mathbf{6}} \otimes \mathbf{6}$. 

For the fully strange $ss\bar s\bar s$ tetraquark systems, there are only two interpolating currents survived since the symmetric flavor structures
\begin{equation}
  \begin{aligned}
     & J_{\alpha \mu \nu}^{s1}=s_a^T C\gamma_{\alpha}  s_b\left(\bar{s}_a \sigma_{\mu \nu} C \bar{s}_b^T\right) - s_a^T C \sigma_{\mu \nu} s_b\left(\bar{s}_a\gamma_{\alpha} C \bar{s}_b^T\right), \\
          & J_{\alpha \mu \nu}^{s3}=s_a^T C\gamma_{\alpha}  s_b\left(\bar{s}_a \sigma_{\mu \nu}\gamma_{5}C \bar{s}_b^T\right) - s_a^T C \sigma_{\mu \nu}\gamma_{5} s_b\left(\bar{s}_a\gamma_{\alpha} C \bar{s}_b^T\right), \\
   \end{aligned}  
\label{Eq:current_ssss}
\end{equation}
in which both of them have the antisymmetric color structure $\mathbf{3} \otimes \bar{\mathbf{3}}$.

To investigate the physical states with definite quantum numbers, we consider the  couplings between the interpolating current and different hadron states as follows
\begin{align}
%\begin{aligned}
\label{Eq:coupling1}
& \left\langle 0\left|J_{\alpha \mu \nu}\right| 0^{(-P) C}(p)\right\rangle=Z_1^0 p_\alpha g_{\mu \nu}+Z_2^0 p_\mu g_{\alpha \nu}+Z_3^0 p_\nu g_{\alpha \mu}+Z_4^0 p_\alpha p_\mu p_\nu  \,,\\ \label{Eq:coupling2}
& \left\langle 0\left|J_{\alpha \mu \nu}\right| 0^{PC}(p)\right\rangle=Z_5^0 \varepsilon_{\alpha \mu \nu \tau} p^\tau  \,,\\ \label{Eq:coupling3}
& \left\langle 0\left|J_{\alpha \mu \nu}\right| 1^{PC}(p)\right\rangle=Z_1^1 \epsilon_\alpha g_{\mu \nu}+Z_2^1 \epsilon_\mu g_{\alpha \nu}+Z_3^1 \epsilon_\nu g_{\alpha \mu}+Z_4^1 \epsilon_\alpha p_\mu p_\nu+Z_5^1 \epsilon_\mu p_\alpha p_\nu+Z_6^1 \epsilon_\nu p_\alpha p_\mu  \,,\\ \label{Eq:coupling4}
& \left\langle 0\left|J_{\alpha \mu \nu}\right| 1^{(-P) C}(p)\right\rangle=Z_7^1 \varepsilon_{\alpha \mu \nu \tau} \epsilon^\tau+Z_8^1 \varepsilon_{\alpha \mu \tau \lambda} \epsilon^\tau p^\lambda p_\nu+Z_9^1 \varepsilon_{\alpha \nu \tau \lambda} \epsilon^\tau p^\lambda p_\mu  \,,\\ \label{Eq:coupling5}
& \left\langle 0\left|J_{\alpha \mu \nu}\right| 2^{(-P) C}(p)\right\rangle=Z_1^2 \epsilon_{\alpha \mu} p_\nu+Z_2^2 \epsilon_{\alpha \nu} p_\mu+Z_3^2 \epsilon_{\mu \nu} p_\alpha  \,,\\ \label{Eq:coupling6}
& \left\langle 0\left|J_{\alpha \mu \nu}\right| 2^{PC}(p)\right\rangle=Z_4^2 \varepsilon_{\alpha \mu \tau \theta} \epsilon_\nu^{~\tau} p^\theta+Z_5^2 \varepsilon_{\alpha \nu \tau \theta} \epsilon_\mu^{~\tau} p^\theta  \,,\\
& \left\langle 0\left|J_{\alpha \mu \nu}\right| 3^{PC}(p)\right\rangle=Z_1^3 \epsilon_{\alpha \mu \nu} \, ,
%\end{aligned}
\label{Eq:coupling7}
\end{align}
where $\epsilon_\alpha, \epsilon_{\alpha\mu}, \epsilon_{\alpha\mu\nu}$ are the polarization tensors for the spin-1, spin-2 and spin-3 states, respectively. 
It should be noted that the interpolating currents in Eq.~(\ref{Eq:current}) and ~(\ref{Eq:current_ssss}) can not couple to any spin-3 state since their last two Lorentz indices are antisymmetric while the spin-3 polarization tensor $\epsilon_{\alpha\mu\nu}$ is completely symmetric.

Since the parities for the currents $J_{\alpha \mu \nu }^{1}$, $J_{\alpha \mu \nu }^{1\prime}$, $J_{\alpha \mu \nu }^{4}$, $J_{\alpha \mu \nu }^{4\prime}$, $J_{\alpha \mu \nu }^{s1}$ and $J_{\alpha \mu \nu }^{2}$, $J_{\alpha \mu \nu }^{2\prime}$, $J_{\alpha \mu \nu }^{3}$, $J_{\alpha \mu \nu }^{3\prime}$, $J_{\alpha \mu \nu }^{s3}$ are opposite, they couple to the $2^{+-}$ tetraquark states via different coupling relations in Eq.~\eqref{Eq:coupling5} and Eq.~\eqref{Eq:coupling6}, respectively. 
For the currents $J_{\alpha \mu \nu }^{2,2\prime,3,3\prime,s3}$, we can rewrite the coupling in another way
\begin{equation}
\begin{aligned}
\left\langle 0\left|J_{\alpha \mu \nu}^{2,2\prime,3,3\prime,s3}\right| 2^{+-}(p)\right\rangle & =Z_4^2 \varepsilon_{\alpha \mu \tau \theta} \epsilon_\nu^{~\tau} p^\theta+Z_5^2 \varepsilon_{\alpha \nu \tau \theta} \epsilon_\mu^{~\tau} p^\theta \\
& =f^{+}\left(\varepsilon_{\alpha \mu \tau \theta} \epsilon_\nu^{~\tau} p^\theta+\varepsilon_{\alpha \nu \tau \theta} \epsilon_\mu^{~\tau}p^\theta\right)+f^{-}\left(\varepsilon_{\alpha \mu \tau \theta} \epsilon_\nu^{~\tau} p^\theta-\varepsilon_{\alpha \nu \tau \theta} \epsilon_\mu^{~\tau} p^\theta\right)\\
& =f^{-}\left(\varepsilon_{\alpha \mu \tau \theta} \epsilon_\nu^{~\tau} p^\theta-\varepsilon_{\alpha \nu \tau \theta} \epsilon_\mu^{~\tau} p^\theta\right)\, ,
\end{aligned}
\label{Eq:coupling6a}
\end{equation}
in which the Lorentz indices $\mu\nu$ are antisymmetric in the last step to be consistent with those in the interpolating currents. 
%Since the indices $\mu$ and  $\nu $ are anti-symmetric in Eq.~(\ref{Eq:current}), the symmetric part in above equation vanishes, in other words, only one $2^{+-}$ physical state can be obtained from the current $J_{\alpha \mu \nu}^{2}$, $J_{\alpha \mu \nu}^{2\prime}$, $J_{\alpha \mu \nu}^{3}$ and $J_{\alpha \mu \nu}^{3\prime}$, respectively. 
One can construct the normalized projector operator for the $2^{+-}$ state 
\begin{equation}
\mathbb{P}(\alpha_{1},\mu_{1},\nu_{1},\alpha_{2},\mu_{2},\nu_{2})=\frac{1}{20}\sum \left(\varepsilon_{\alpha_{1} \mu_{1}  \tau_{1}  \theta_{1} } \epsilon_{\nu_{1} }^{~\tau_{1} } p^{\theta _{1}}-\varepsilon_{\alpha_{1}  \nu_{1}  \tau_{1}  \theta_{1} } \epsilon_{\mu_{1} }^{~\tau_{1} } p^{\theta_{1} }\right)
\left(\varepsilon_{\alpha_{2}  \mu_{2} \tau_{2} \theta_{2}} \epsilon_{\nu_{2}}^{~\tau_{2}*} p^{\theta_{2}}-\varepsilon_{\alpha_{2} \nu_{2} \tau_{2} \theta_{2}} \epsilon_{\mu_{2}}^{~\tau_{2}*} p^{\theta_{2}}\right) / p^2  \,,
\label{Eq:Projector1}
\end{equation}
where the summation over polarization of the tensor $\epsilon_{\alpha \beta}$ is
\begin{equation}
\sum \epsilon_{\alpha_{1} \beta_{1}} \epsilon_{\alpha_{2} \beta_{2}}^{*}=\frac{1}{2}(\eta_{\alpha_{1} \alpha_{2}} \eta_{\beta_{1} \beta_{2}}+\eta_{\alpha_{1} \beta_{2}} \eta_{\beta_{1} \alpha_{2}}-\frac{2}{3} \eta_{\alpha_{1} \beta_{1}} \eta_{\alpha_{2} \beta_{2}}) \,,
\end{equation}
with
\begin{equation}
\eta_{\alpha\beta}=\frac{p_{\alpha} p_{\beta}}{p^{2}}-g_{\alpha \beta} \,.
\end{equation}
One may wonder that there should be a coupling corresponding to the Lorentz structure $\varepsilon_{\mu \nu \tau \theta} \epsilon_\alpha^\tau p^\theta $ in Eq.~\eqref{Eq:coupling6}. As a matter of fact, the projector constructed by this structure is the same as that from the antisymmetric part in Eq.~\eqref{Eq:coupling6a}, so that there are only two independent tensor structures in Eq.~\eqref{Eq:coupling6a}. 

For the currents $J_{\alpha \mu \nu }^{1,1\prime,4,4\prime,s1}$, the coupling relation in  Eq.~\eqref{Eq:coupling5} can be rewritten as
\begin{equation}
  \begin{aligned}
\left\langle 0\left|J_{\alpha \mu \nu}^{1,1\prime,4,4\prime,s1}\right| 2^{+-}(p)\right\rangle &=Z_1^2 \epsilon_{\alpha \mu} p_\nu+Z_2^2 \epsilon_{\alpha \nu} p_\mu+Z_3^2 \epsilon_{\mu \nu} p_\alpha \\
&=Z_1^2 \epsilon_{\alpha \mu} p_\nu+Z_2^2 \epsilon_{\alpha \nu} p_\mu\\
&=
%f_{+}^{\prime} (\epsilon_{\alpha \mu } p_\nu   +\epsilon_{\alpha \nu} p_\mu)+
f^{-\prime}(\epsilon_{\alpha \mu } p_\nu  -\epsilon_{\alpha \nu} p_\mu)\, .
\end{aligned}
\label{Eq:coupling5a}
\end{equation}
with the normalized projector operator
\begin{equation}
  \mathbb{P}^{\prime}(\alpha_{1},\mu_{1},\nu_{1},\alpha_{2},\mu_{2},\nu_{2})=\frac{1}{20}\sum \left( \epsilon_{\alpha _{1} \mu_{1} } p_{\nu_{1}}-\epsilon_{\alpha _{1} \nu _{1} } p_{\mu _{1} }\right)
  \left( \epsilon_{\alpha _{2} \mu_{2} } p_{\nu_{2}}-\epsilon_{\alpha _{2} \nu _{2} } p_{\mu _{2} }\right) / p^2  \, .
  \label{Eq:Projector2}
\end{equation}
In our calculations, we find that the $2^{+-}$ tetraquark states extracted from the currents $J_{\alpha \mu \nu}^{1, 1\prime, 2, 2\prime,s1}$ are the same with those from $J_{\alpha \mu \nu}^{3, 3\prime, 4, 4\prime,s3}$, respectively. In the following analyses, we consider only the interpolating currents $J_{\alpha \mu \nu }^{1}$, $J_{\alpha \mu \nu }^{1\prime}$, $J_{\alpha \mu \nu }^{2}$, $J_{\alpha \mu \nu }^{2\prime}$ and $J_{\alpha \mu \nu }^{s1}$ to investigate the mass spectra of the $2^{+-}$ tetraquark states.

\section{Formalism of QCD sum rules}\label{Sec:3}
The two-point correlation function of the current in Eq.~(\ref{Eq:current}) can be written as
\begin{equation}
\begin{aligned}
 \Pi_{\alpha_{1}\mu_{1} \nu_{1},\alpha_{2}\mu_{2} \nu_{2}}(p^{2}) &=i \int d^{4} x e^{i p \cdot x}\left\langle 0\left|T\left[J_{\alpha_{1}\mu_{1} \nu_{1}}(x) J_{\alpha_{2}\mu_{2} \nu_{2}}^{\dagger}(0)\right]\right| 0\right\rangle \, .
 \label{Eq:correlator}
\end{aligned}
\end{equation}
We shall investigate only the $J^{PC}=2^{+-}$ tetraquark states in this work, which can be extracted by applying the projection operator defined in Eq.~\eqref{Eq:Projector2}
\begin{equation}
  \begin{aligned}
    \Pi_{2}(p^{2}) &=\mathbb{P}^{(\prime)}(\alpha_{1},\mu_{1},\nu_{1},\alpha_{2},\mu_{2},\nu_{2}) \Pi_{\alpha_{1}\mu_{1} \nu_{1},\alpha_{2}\mu_{2} \nu_{2}}(p^{2}) \, ,
   \label{Eq:Pi_2}
  \end{aligned}
  \end{equation}

At the hadronic level, the correlation function $\Pi_{2}(p^{2})$ can be usually described via the dispersion relation
\begin{equation}
\Pi_{2}(p^{2})=\frac{(p^{2})^{N}}{\pi} \int_{0}^{\infty} \frac{\operatorname{Im} \Pi_{2}(s)}{s^{N}\left(s-p^{2}-i \epsilon\right)} d s+\sum_{n=0}^{N-1} b_{n}(p^{2})^{n}\, ,
\label{Cor-Spe}
\end{equation}
where the $b_n$ is the subtraction constant. In QCD sum rules, the imaginary part of the correlation function is defined as the spectral function
\begin{equation}
\rho (s)\equiv\frac{1}{\pi} \text{Im}\Pi_{2}(s)=f^{2}m_{H}^{2}\delta(s-m_{H}^{2})+\text{QCD continuum and higher states}\, ,
\end{equation}
in which the “one pole plus continuum” parametrization assumption is used. The parameters $f$ and $m_{H}$ are the coupling constant and mass of the lowest-lying hadron state $H$, respectively.

To improve the convergence of the OPE series and suppress the contributions from continuum and higher states, Borel transformation can be performed to the correlation functions in both hadron and quark-gluon levels. The QCD sum rules are then obtained as
\begin{equation}
\Pi_{2}\left(s_{0}, M_{B}^{2}\right)=f^{2} m_{H}^{2}e^{-m_{H}^{2} / M_{B}^{2}}=\int_{0}^{s_{0}} d s e^{-s / M_{B}^{2}} \rho(s)\, ,
\end{equation}
where $M_B$ is the Borel mass introduced via the Borel transformation and $s_0$ is the continuum threshold. Since the Borel mass is an intermediate parameter, it should not be relevant to the physical state. These two parameters can be determined by requiring a suitable OPE convergence and a big enough pole contribution in the QCD sum rule analyses.
Then the hadron mass of the lowest-lying tetraquark state can be extracted as
\begin{equation}
   \begin{aligned}
   m_{H}\left(s_0, M_B^2\right) & =\sqrt{\frac{\frac{\partial}{\partial\left(-1 / M_B^2\right)} \Pi_{2}\left(s_0, M_B^2\right)}{\Pi_{2}\left(s_0, M_B^2\right)}} \,.
  \end{aligned}
\end{equation}

We evaluate the correlation functions for the light tetraquark states with $J^{PC}=2^{+-}$ up to dimension 10 condensates. The contribution from operators with higher dimensions are extremely small and we will not take them into consideration in this work. We list the results in the Appendix since these expressions are relatively complicated. For the nonstrange $ud\bar{u}\bar{d}$ systems, we have neglected the masses of light quarks in the chiral limit so that there is no contribution from odd dimensional condensates, such as the quark condensates and quark-gluon mixed condensates. For the $us\bar{u}\bar{s}$ and $ss\bar{s}\bar{s}$ systems, the strange quark mass will be taken into consideration. For the high dimension condensates, we use the factorization assumptions and set the factors to be one in our analyses.

\section{Numerical analyses and mass predictions}\label{Sec:4}
In this section, we perform the QCD sum rule analyses for the light exotic tetraquark states with $J^{PC}=2^{+-}$. We use the following values for various QCD parameters~\cite{Jamin:2002ev,Narison:2011xe,Narison:2018dcr,ParticleDataGroup:2022pth}.
\begin{equation}
  \begin{aligned}
  m_{u} & = m_d=m_q=0\, ,\\
  m_{s} & =93_{-5}^{+11}  ~\mathrm{MeV}\, ,\\
  \dqq & =-(0.24 \pm 0.01)^3 ~\mathrm{GeV}^3 \,, \\
  \dss & =(0.8\pm 0.1)\times \dqq \,, \\
  \dqGq & =-(0.8 \pm 0.2) \times \dqq ~\mathrm{GeV}^2 \,, \\
  \dsGs & =(0.8\pm 0.2) \times \dqGq\, ,\\ 
  \dGG & =(6.35 \pm 0.35) \times 10^{-2} ~\mathrm{GeV}^4 \,.
  %\dGGG & =-(8.2 \pm 1.0) \times \dGG ~\mathrm{GeV}^2 \, .
  \end{aligned}
  \end{equation}
The above condensate values are taken at energy renormalization scale $\mu=1\,\mathrm{GeV}$ while the $s$ quark mass is taken at $\mu=2\,\mathrm{GeV}$. To maintain energy renormalization scale consistency, we use renormalization group to run the $s$ quark mass at $\mu=1\,\mathrm{GeV}$, as  adopted in Ref.~\cite{Wang:2019nln}.

\begin{figure}[t!!]
  \centering
  \includegraphics[width=8cm]{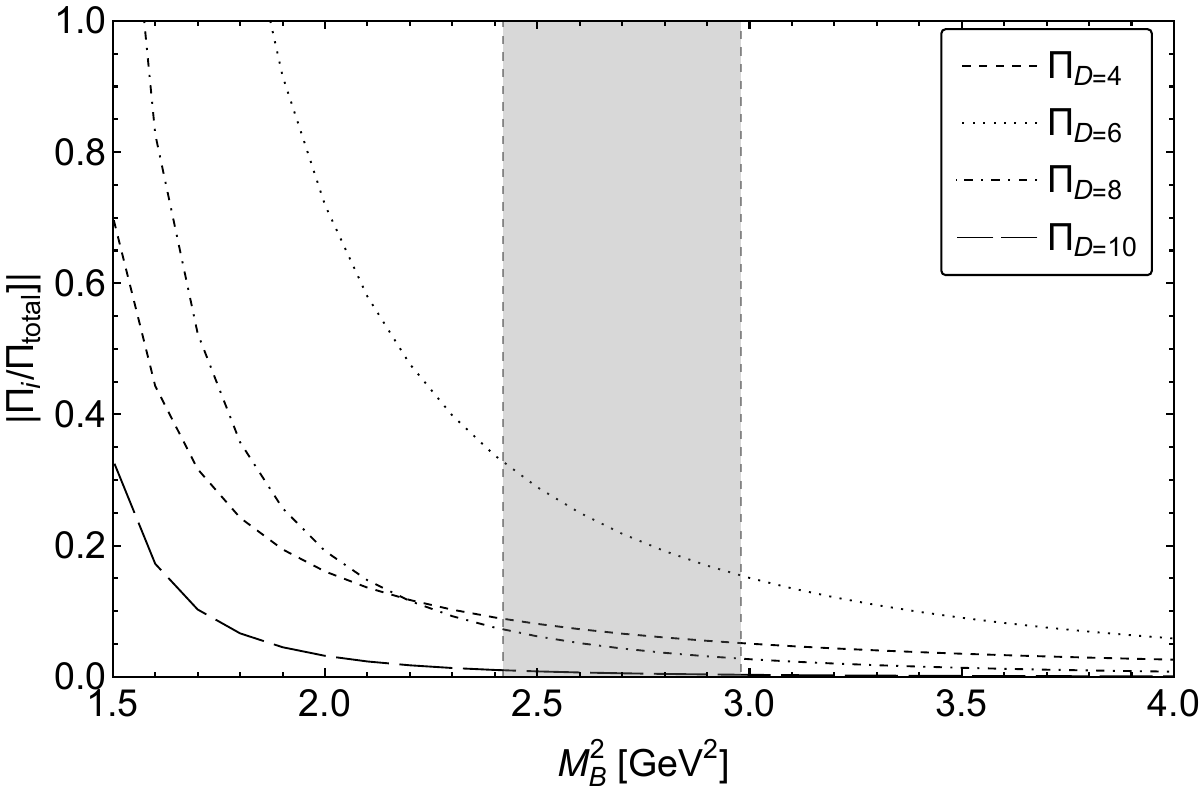}\\
  \caption{OPE convergence for the $ud\bar{u}\bar{d}$ tetraquark state with $J^{PC}=2^{+-}$ extracted from the current $J_{\alpha\mu\nu}^{1}$.}
  \label{fig:Convergence}
\end{figure}

\begin{figure}[t!!]
  \centering
  \subfigure[]{\includegraphics[width=8cm]{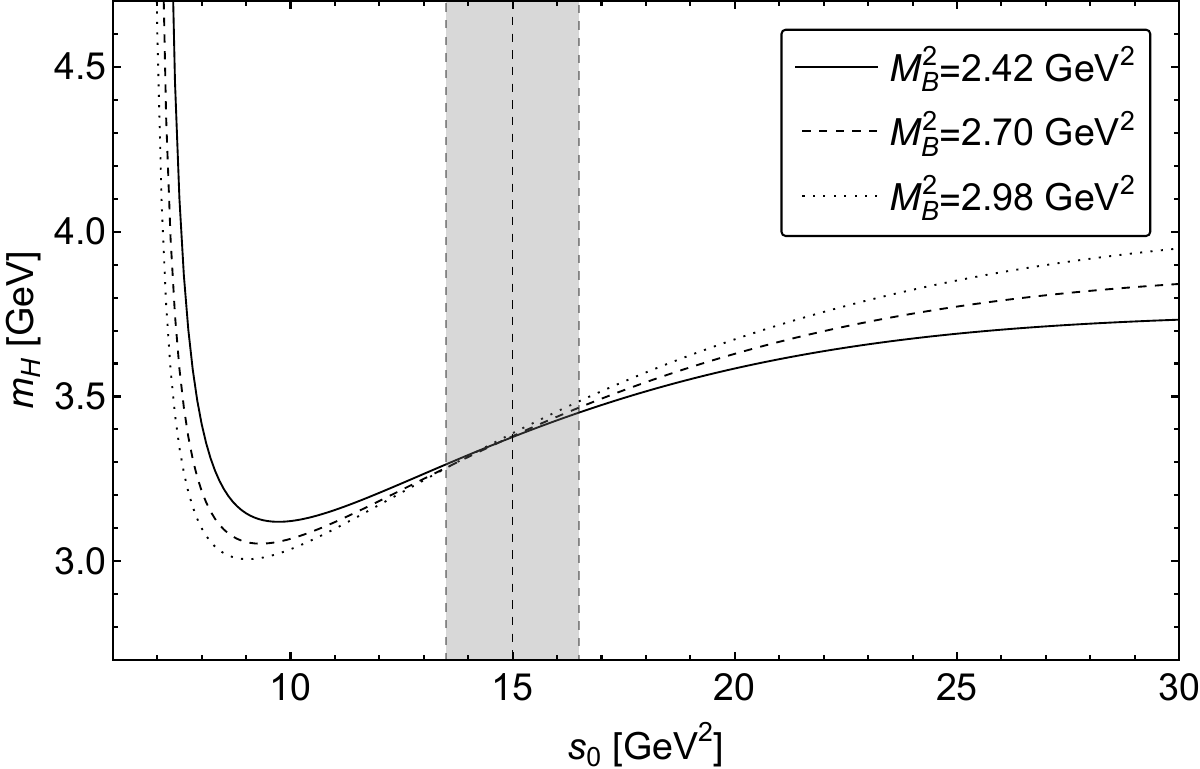}\label{fig:s0-mH}}
  \subfigure[]{\includegraphics[width=8.1cm]{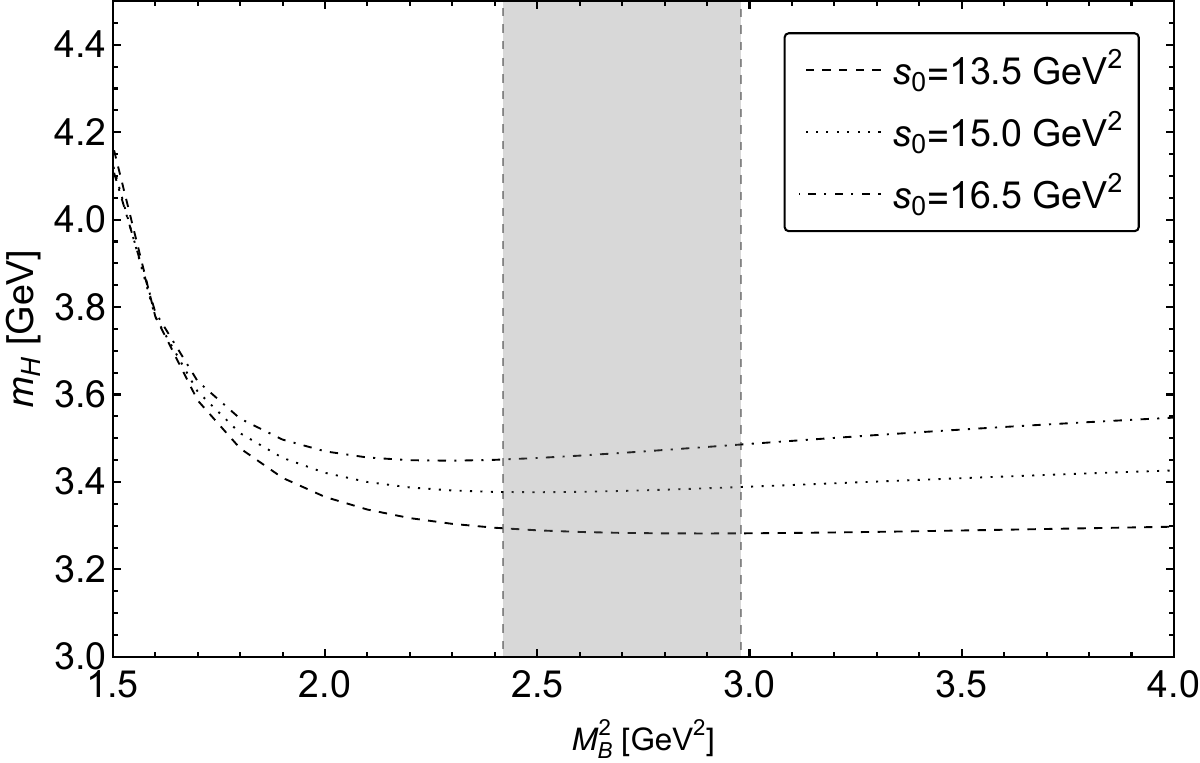}\label{fig:MB-mH}}\\
  \caption{Variation of $m_{H}$ with $s_{0}$ and $M_{B}^{2}$ corresponding to the $J^{PC}=2^{+-}$ $ud\bar{u}\bar{d}$ tetraquark state extracted from current $J_{\alpha\mu\nu}^{1}$.}
  \label{fig:Result}
\end{figure}

We take $J_{\alpha\mu\nu}^{1}$ as an example to show the details of our numerical analyses for the nonstrange $ud\bar{u}\bar{d}$ tetraquark state. To extract the output parameters, the Borel parameter $M_{B}^{2}$ should be large enough to guarantee the convergence of OPE series. We require the contribution from the high dimension $D>8$ condensates to be less than 1\%, i.e
\begin{equation}
R_{D>8}=\left|\frac{\Pi^{D>8}\left(M_{B}^{2}, \infty\right)}{\Pi^{t o t}\left(M_{B}^{2}, \infty\right)}\right|<1 \% \,.
\end{equation}
This requirement leads to the lower bound on the Borel parameter $M_{B}^{2}\geq 2.42~\mathrm{GeV}^{2}$.
We show the contribution ratios from various condensates in Fig.~\ref{fig:Convergence}, from which one finds that the convergence of OPE series can be well ensured. In this $ud\bar{u}\bar{d}$ tetraquark system, the dominant nonperturbative effect comes from the dimension 6 four-quark condensate $\dqq^2$, since the contributions from the quark condensate $\dqq$ and quark-gluon mixed condensate $\dqGq$ vanish in the chiral limit. 
To get the upper bound on $M_{B}^{2}$, we need to fix the value of $s_{0}$ at first. As mentioned in Sec.~\ref{Sec:3}, the hadron mass $m_{H}$ should be irrelevant to the intermediate parameter $M_{B}^{2}$. In Fig.~\ref{fig:s0-mH}, we show the variations of $m_{H}$ with respect to the continuum threshold $s_{0}$ for various Borel parameter $M_{B}^{2}$. It is shown that the variation of $m_{H}$ with $M_{B}^{2}$ can be minimized in the parameter working region $13.5\leq s_0\leq 16.5~\mathrm{GeV}^{2}$. Then the upper bound on $M_{B}^{2}$ can be determined by requiring the following pole contribution be larger than 50\%
\begin{equation}
\text{Pole Contribution}=\frac{\Pi\left(M_{B}^{2}, s_0\right)}{\Pi\left(M_{B}^{2}, \infty\right)}>50 \% \, .
\end{equation}
Finally, the working region of the Borel parameter can be determined to be $2.42\leq M_{B}^{2}\leq 2.98~\mathrm{GeV}^{2}$. We show the Borel curves in the above parameter working regions in Fig.~\ref{fig:MB-mH}, in which the QCD sum rules are stable enough to predict the hadron mass of $ud\bar{u}\bar{d}$ tetraquark state as
\begin{equation}
  \begin{aligned}
    m_{ud\bar{u}\bar{d}}^{1}&=3.38_{-0.12}^{+0.13}~\mathrm{GeV}\, .
  \end{aligned}
\end{equation}

The errors are mainly from the uncertainties of the continuum threshold $s_{0}$, various QCD condensates $\dqq$, $\dGG$ and $\dqGq$. The error from the Borel mass is small enough to be neglected.

\begin{figure}[t!!]
  \centering
  \includegraphics[width=8cm]{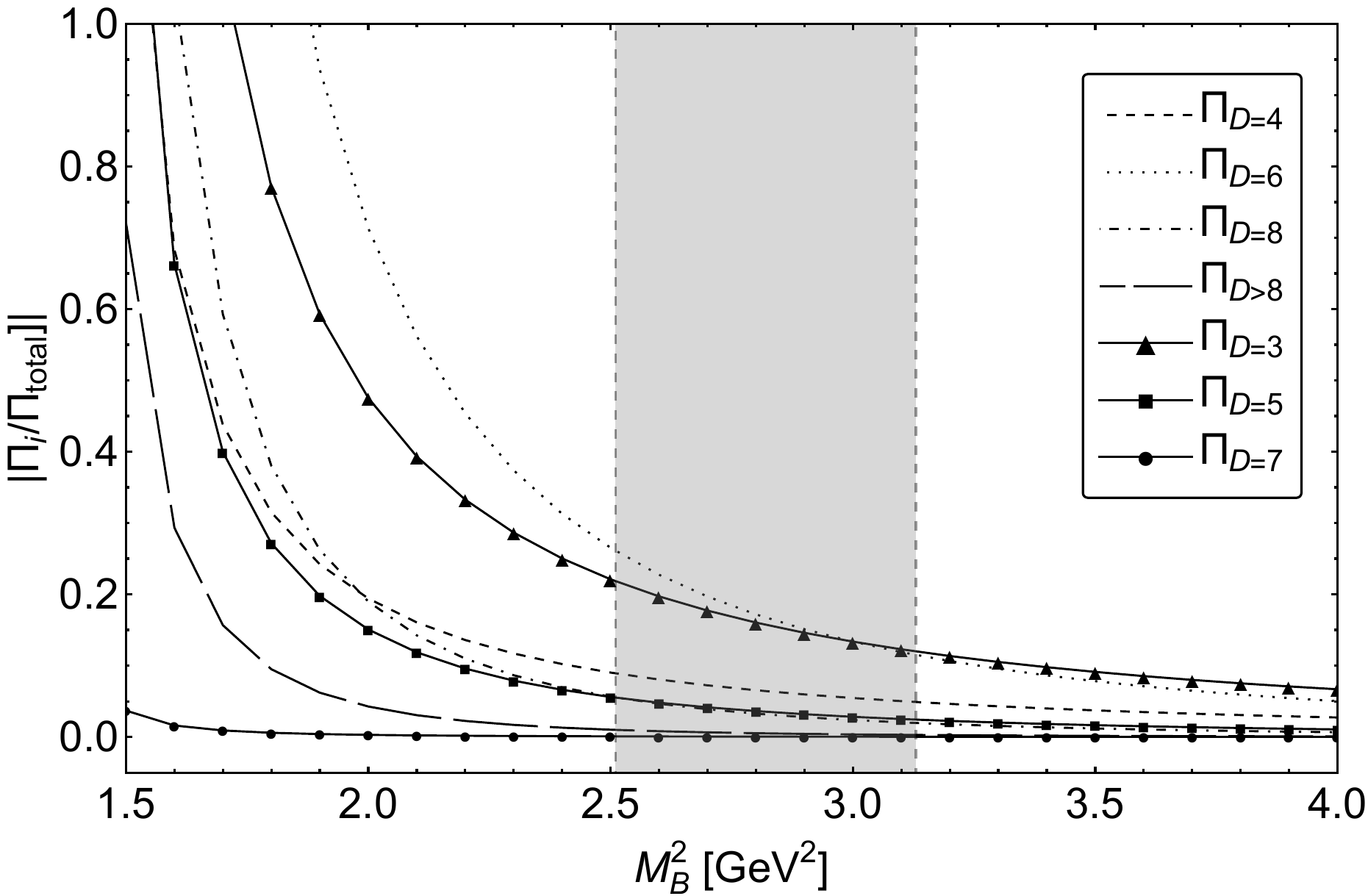}\\
  \caption{OPE convergence for the $us\bar{u}\bar{s}$ tetraquark state with $J^{PC}=2^{+-}$ extracted from the current $J_{\alpha\mu\nu}^{1}$.}
  \label{fig:Convergence-s}
\end{figure}

Replacing $d$ to $s$ quark field, one can perform similar QCD sum rule calculations and analyses for the hidden-strange $us\bar{u}\bar{s}$ tetraquark systems. As mentioned in Sec.~\ref{Sec:4}, the correlation functions for  $us\bar{u}\bar{s}$ system contain the contributions from the dimension 3 quark condensates and dimension 5 quark-gluon mixed condensates. In Fig.~\ref{fig:Convergence-s}, we show the OPE convergence for the $us\bar{u}\bar{s}$ tetraquark state from the current $J_{\alpha\mu\nu}^{1}$, from which one finds that the dominant nonperturbative effect is still from the four-quark condensates. However, the contribution from the quark condensates is significant and even larger than the four-quark condensates for the big values of $M_B^2$.  This is very different from the situation in the $ud\bar{u}\bar{d}$ system, where the quark condensates and quark-gluon mixed condensates give no contribution to the correlation function. 

\begin{figure}[t!!]
  \centering
  \subfigure[]{\includegraphics[width=8cm]{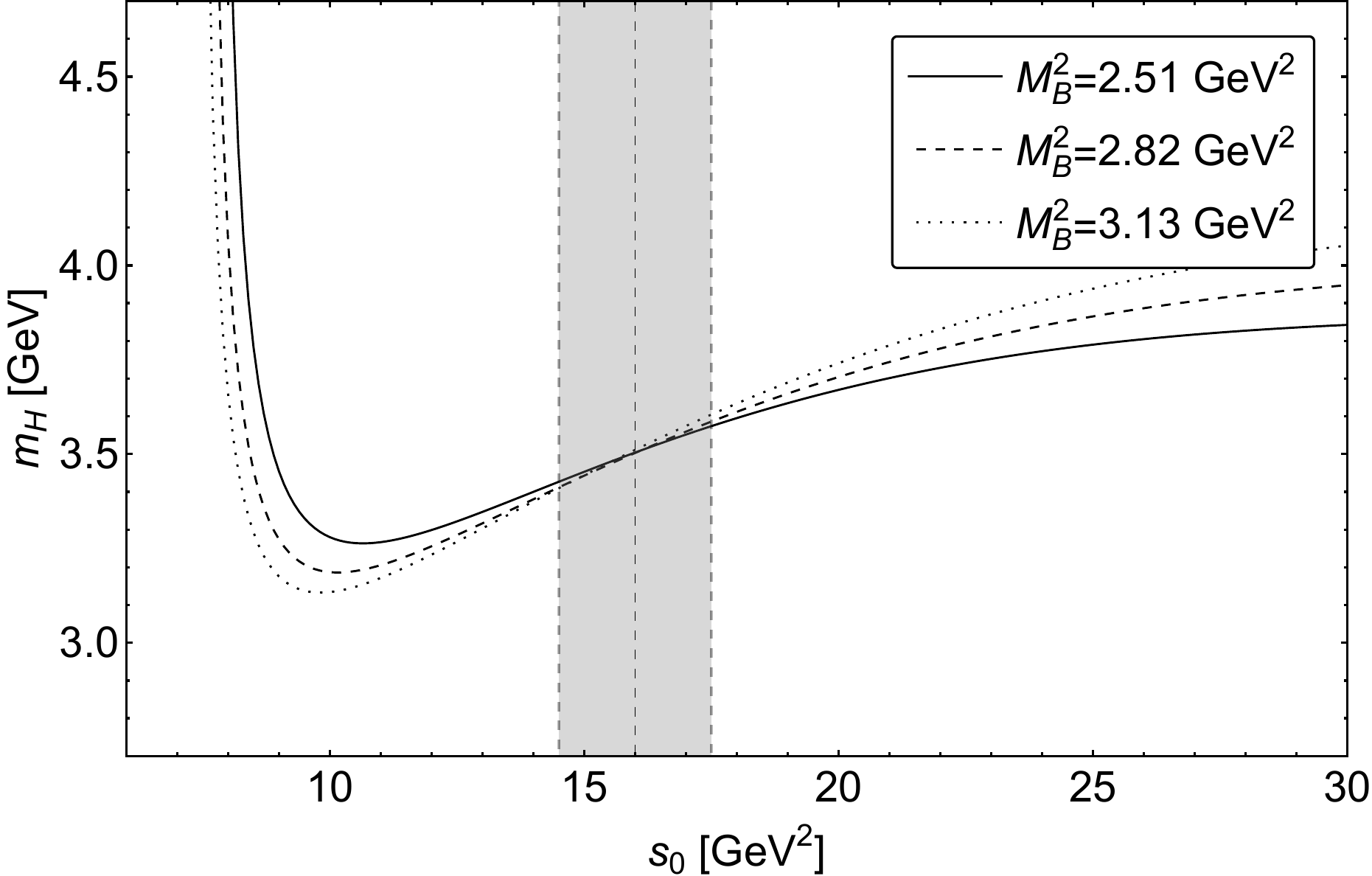}\label{fig:s0-mH-s}}
  \subfigure[]{\includegraphics[width=8.1cm]{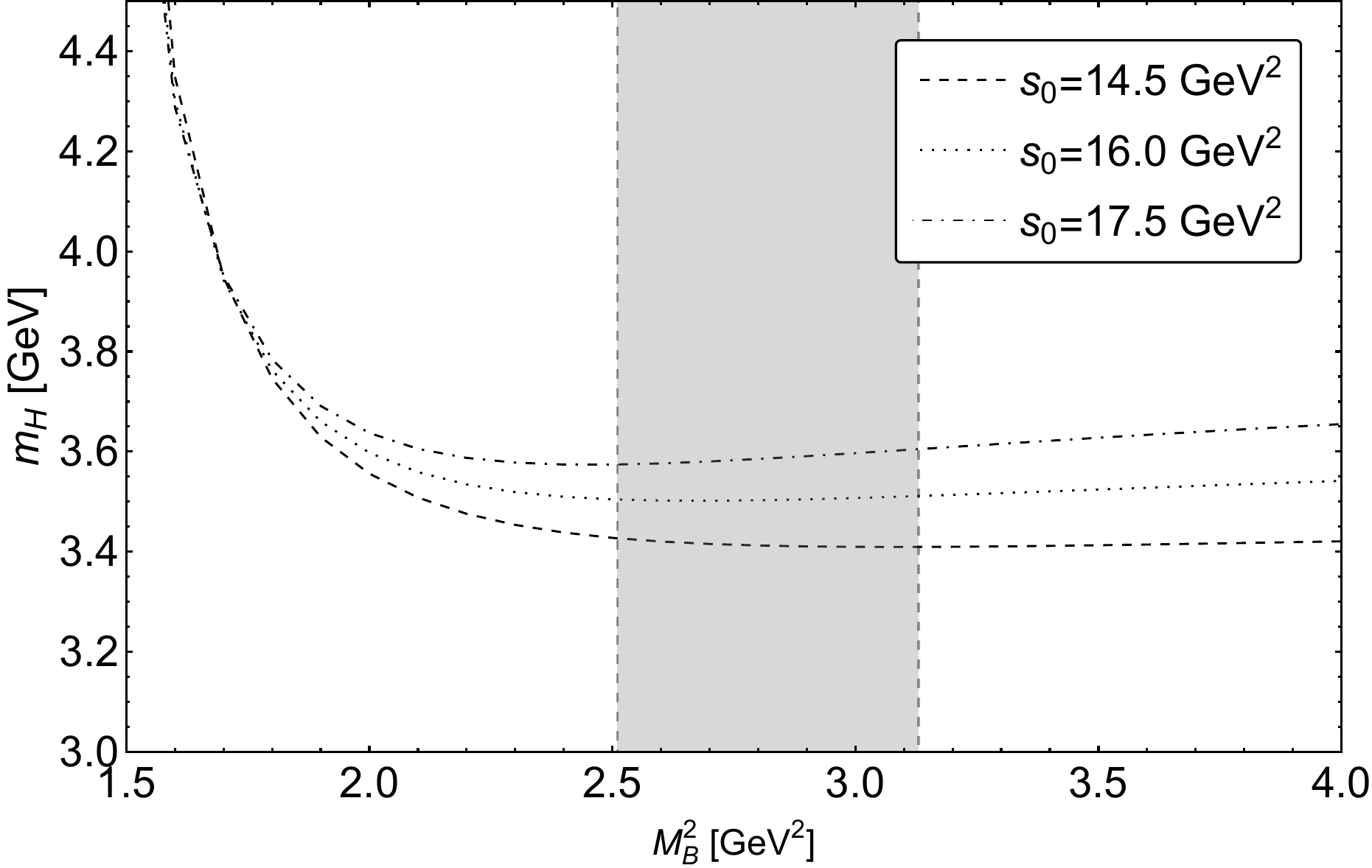}\label{fig:MB-mH-s}}\\
  \caption{Variation of $m_{H}$ with $s_{0}$ and $M_{B}^{2}$ corresponding to the $J^{PC}=2^{+-}$ $us\bar{u}\bar{s}$ tetraquark state extracted from current $J_{\alpha\mu\nu}^{1}$.}
  \label{fig:Result-s}
\end{figure}

For the $us\bar{u}\bar{s}$ system with $J_{\alpha\mu\nu}^{1}$, the parameter working regions can be obtained as $2.51 \leq M_{B}^{2}\leq 3.13~\mathrm{GeV}^{2}$ and $14.5\leq s_0\leq 17.5~\mathrm{GeV}^{2}$ after similar numerical analyses, where the Borel window is slightly broader than that for the $ud\bar{u}\bar{d}$ system. Then the hadron mass can be predicted as
\begin{equation}
  \begin{aligned}
    m_{us\bar{u}\bar{s}}^{1}&=3.50_{-0.12}^{+0.13}~\mathrm{GeV}\, ,
  \end{aligned}
\end{equation}
which is about 100 MeV higher than the $ud\bar{u}\bar{d}$ tetraquark state. We show the corresponding mass curves in Fig.~\ref{fig:Result-s}. 

\begin{figure}[t!!]
  \centering
  \includegraphics[width=8cm]{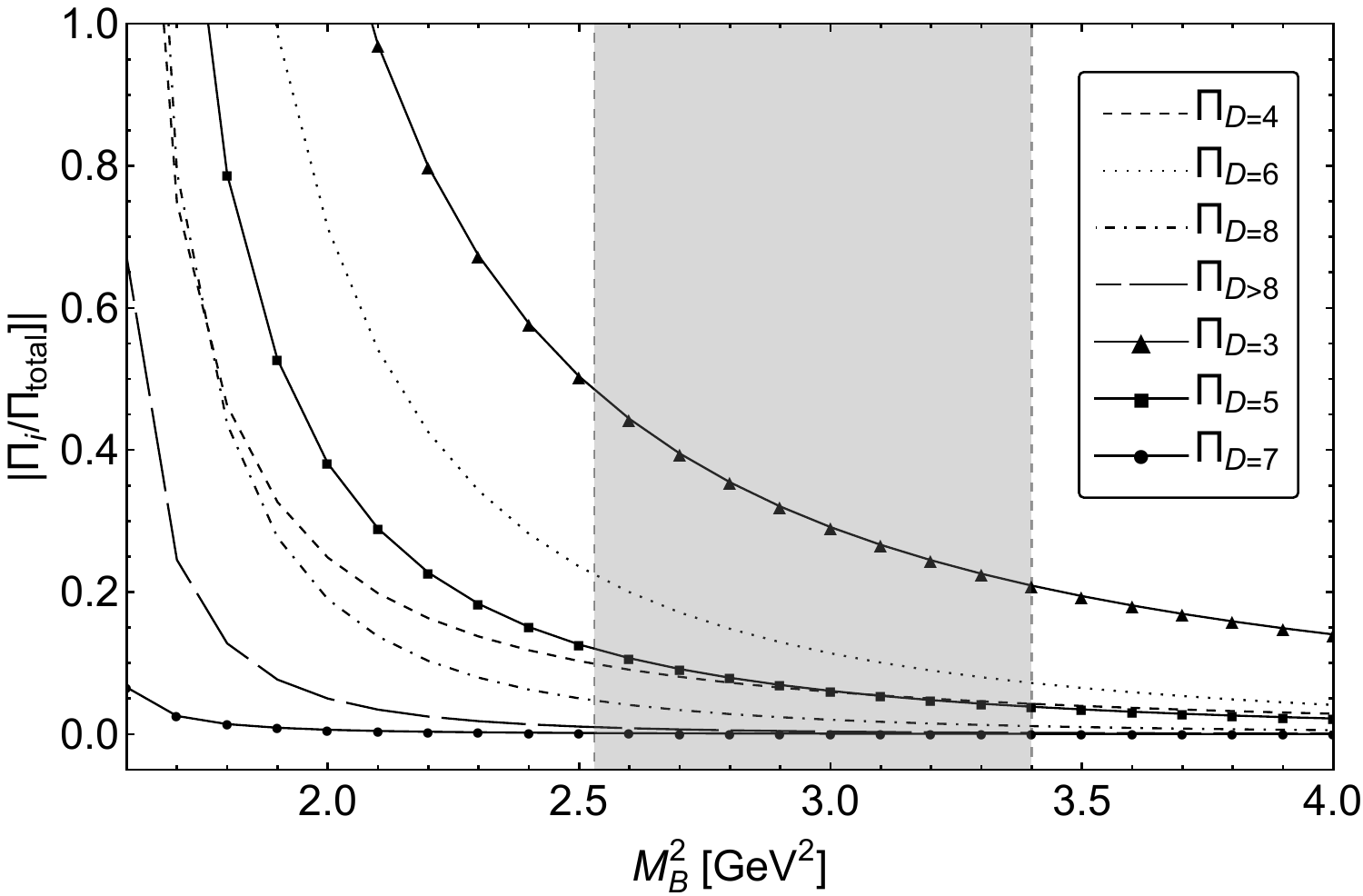}\\
  \caption{OPE convergence for the $ss\bar{s}\bar{s}$ tetraquark state with $J^{PC}=2^{+-}$ extracted from the current $J_{\alpha\mu\nu}^{1}$.}
  \label{fig:Convergence-ss}
\end{figure}

\begin{figure}[t!!]
  \centering
  \subfigure[]{\includegraphics[width=8cm]{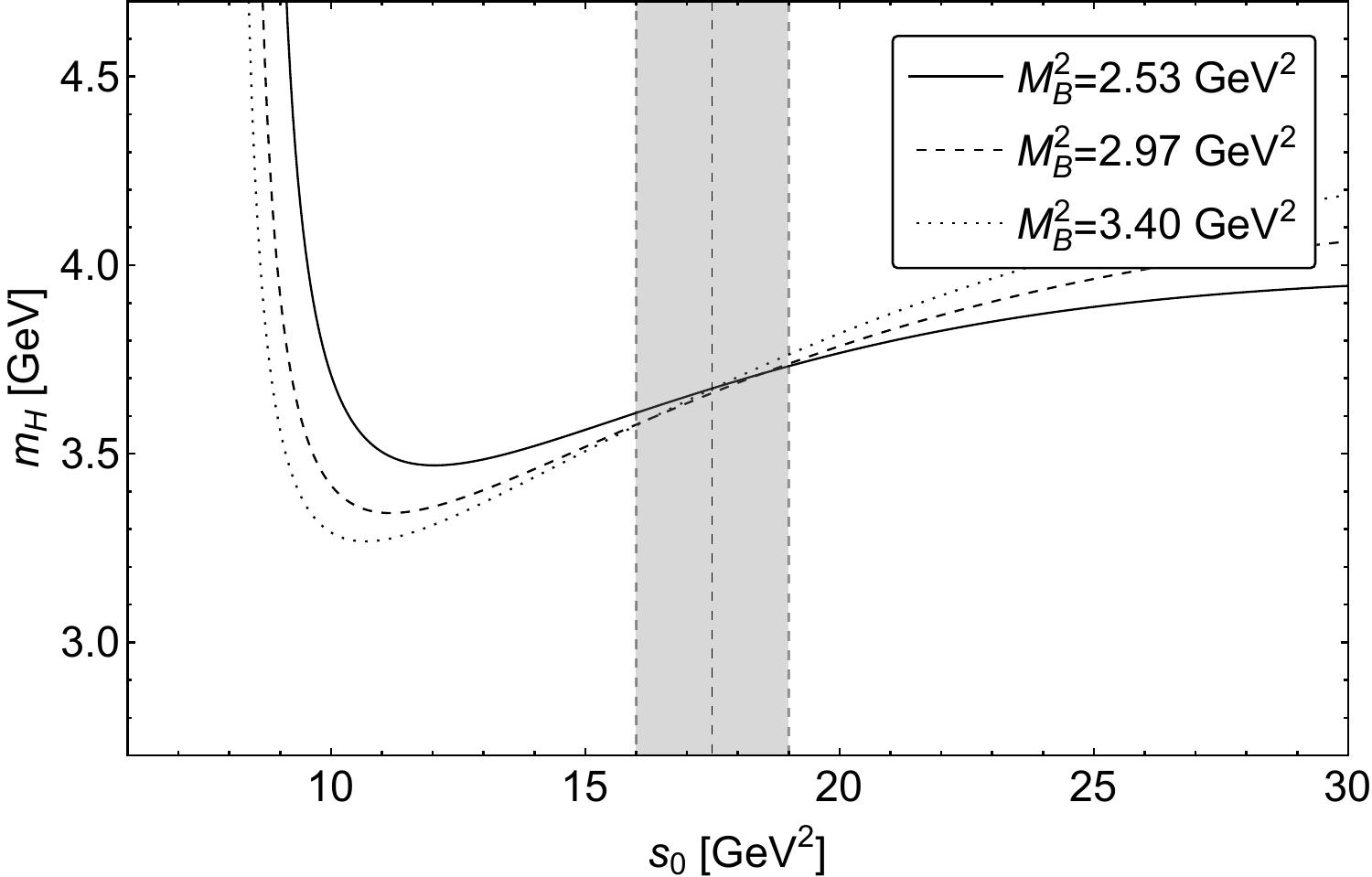}\label{fig:s0-mH-ss}}
  \subfigure[]{\includegraphics[width=8.1cm]{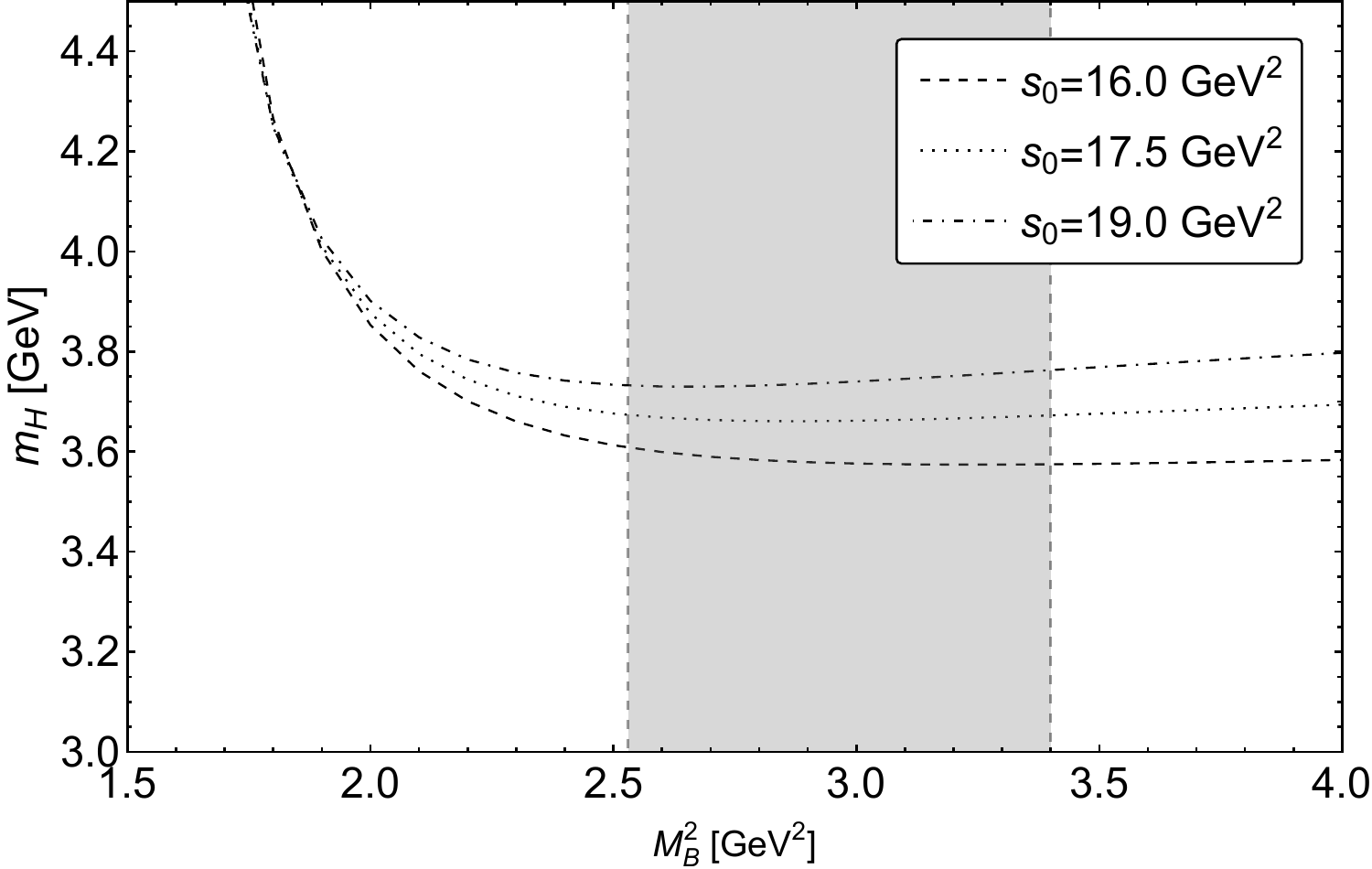}\label{fig:MB-mH-ss}}
  \\
  \caption{Variation of $m_{H}$ with $s_{0}$ and $M_{B}^{2}$ corresponding to the $J^{PC}=2^{+-}$ $ss\bar{s}\bar{s}$ tetraquark state extracted from current $J_{\alpha\mu\nu}^{s1}$.}
  \label{fig:Result-ss}
\end{figure}

For the fully strange system, we show the OPE convergence for the $ss\bar{s}\bar{s}$ tetraquark state from the current $J_{\alpha\mu\nu}^{s1}$  in Fig.~\ref{fig:Convergence-ss}, from which one finds that the dominant nonperturbative effect is from the dimension 3 quark condensate rather than from the four-quark condensates, and the dimension 5 quark-gluon mixed condensate also plays an important role in the numerical analysis. For this current, the parameter working regions can be obtained as $2.53 \leq M_{B}^{2}\leq 3.40~\mathrm{GeV}^{2}$ and $16.0\leq s_0\leq 19.0~\mathrm{GeV}^{2}$. After similar numerical analyses, the hadron mass can be predicted as
\begin{equation}
  \begin{aligned}
    m_{ss\bar{s}\bar{s}}&=3.66_{-0.09}^{+0.10}~\mathrm{GeV}\, ,
  \end{aligned}
\end{equation}
which is about 100 MeV higher than the $us\bar{u}\bar{s}$ tetraquark state. We show the corresponding mass curves in Fig.~\ref{fig:Result-ss}. For all interpolating currents in Eq.~\eqref{Eq:current} and  Eq.~\eqref{Eq:current_ssss}, we collect the numerical results for the $ud\bar{u}\bar{d}$, $us\bar{u}\bar{s}$ and  $ss\bar{s}\bar{s}$ tetraquark states with $J^{PC}=2^{+-}$ in Table~\ref{Tab:Results}.

%%%%%%%%%%%%%%%%%%%
\begin{table*}[h!]
  \caption{Predicted tetraquark masses and the corresponding parameters for all interpolating currents. The pole contributions are  evaluated at the central values of $s_{0}$ and $M_{B}^{2}$}.
  \renewcommand\arraystretch{1.6} 
  \setlength{\tabcolsep}{1.8em}{   \begin{tabular}{cccccc}
   \hline  \hline 
    &Current    & $s_{0}(\mathrm{GeV}^{2})$    &$M_{B}^{2}(\mathrm{GeV}^{2})$ & Mass$(\mathrm{GeV})$ & Pole Contribution($\%$)\\ \hline
    \multirow{4}{*}{$ud\bar{u}\bar{d}$ }     
    & $J_{\alpha \mu \nu}^{1}$  & $15.0\pm 1.5$  &$2.42 -  2.98$  & $3.38_{-0.12}^{+0.13}$ & $57.9$\\  
    & $J_{\alpha \mu \nu}^{1\prime}$  & $16.5\pm 1.5$  &$2.21 - 3.33$  &$3.51_{-0.10}^{+0.11}$ & $64.7$\\
    & $J_{\alpha \mu \nu}^{2}$ & $15.0\pm 1.5$  &$2.36 -  2.98$  & $3.39_{-0.12}^{+0.13}$ & $58.6$\\ 
    & $J_{\alpha \mu \nu}^{2\prime}$  & $14.5\pm 1.5$  &$2.45 -  2.87$  & $3.34_{-0.12}^{+0.14}$ & $55.8$\\ 
                                             \cline{2-6}
   \multirow{4}{*}{$us\bar{u}\bar{s}$ }     
    & $J_{\alpha \mu \nu}^{1}$  & $16.0\pm 1.5$  &$2.51 -  3.13$  & $3.50_{-0.12}^{+0.13}$ & $58.4$\\  
    & $J_{\alpha \mu \nu}^{1\prime}$  & $18.0\pm 1.5$  &$2.36 - 3.60$  &$3.66_{-0.10}^{+0.11}$ & $65.1$\\
    & $J_{\alpha \mu \nu}^{2}$ & $16.5\pm 1.5$  &$2.46 -  3.25$  & $3.55_{-0.11}^{+0.12}$ & $60.3$\\ 
    & $J_{\alpha \mu \nu}^{2\prime}$  & $16.0\pm 1.5$  &$2.53 - 3.14$  & $3.50_{-0.12}^{+0.12}$ & $58.2$\\ 
       \cline{2-6}
 $ss\bar{s}\bar{s}$  & $J_{\alpha \mu \nu}^{s1}$  & $17.5\pm 1.5$  &$2.53 -  3.40$  & $3.66_{-0.09}^{+0.10}$ & $60.9$\\  
       \hline\hline  
  \label{Tab:Results}
  \end{tabular}}
\end{table*}

\section{Conclusion and Discussion}\label{Sec:5}
We have investigated the mass spectra of the light tetraquark states $ud\bar{u}\bar{d}$, $us\bar{u}\bar{s}$ and $ss\bar{s}\bar{s}$ with exotic quantum number $J^{PC}=2^{+-}$ in QCD sum rules by constructing the interpolating currents with three Lorentz indices. We evaluate the correlation functions and spectral functions up to dimension ten condensates. Our results show that the most important nonperturbative contributions come from the dimension 6 four-quark condensates and dimension 3 quark condensate for the $ud\bar{u}\bar{d}$ and $ss\bar{s}\bar{s}$ tetraquark systems, respectively. For the $us\bar{u}\bar{s}$ system, the contributions from the quark condensates and four-quark condensates are comparable. 

The isospin can be $I=0, 1, 2$ for the nonstrange $ud\bar{u}\bar{d}$ system, $I=0, 1$ for the $us\bar{u}\bar{s}$ system, and $I=0$ for the fully strange $ss\bar{s}\bar{s}$ system. In the SU(2) symmetry, we don't differentiate the up and down quarks in our calculations so that the states in the same tetraquark system with different isospins are degenerate. The predicted hadron masses for the $ud\bar{u}\bar{d}$, $us\bar{u}\bar{s}$ and $ss\bar{s}\bar{s}$ tetraquark states with $J^{PC}=2^{+-}$ are about $3.3 - 3.5\, \mathrm{GeV}$,  $3.5 - 3.7\, \mathrm{GeV}$ and $3.67\, \mathrm{GeV}$, respectively.

\begin{figure}[h!]
	\centering
	\subfigure[]{\includegraphics[width=3cm,height=3cm]{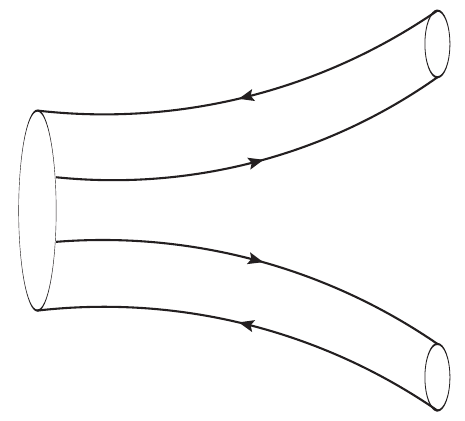}
  \label{fig:decay-1}}\hspace{2cm}
  \subfigure[]{\includegraphics[width=3cm,height=3cm]{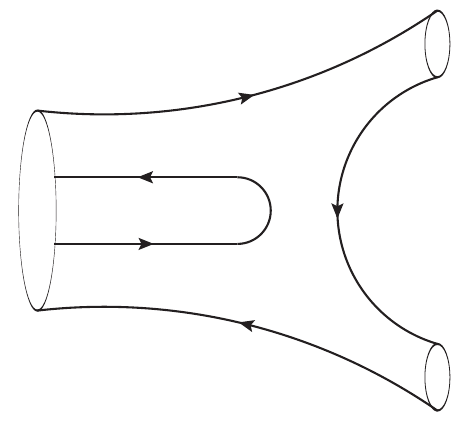}
  \label{fig:decay-2}}\hspace{2cm}
	\subfigure[]{\includegraphics[width=3cm,height=3cm]{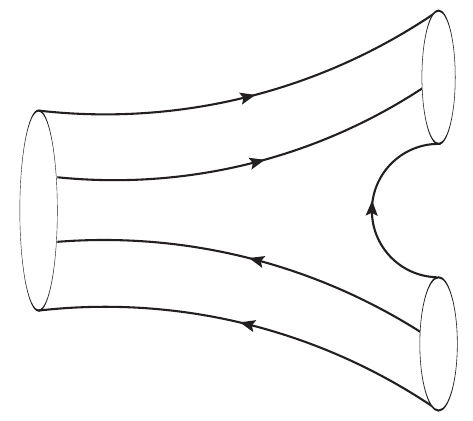}
  \label{fig:decay-3}}
  \\
  \caption{Three possible strong decay mechanisms of the $2^{+-}$ light tetraquark states.}
\end{figure}

The $2^{+-}$ tetraquarks can decay into the two-meson final states via the strong interaction in the spontaneous dissociation mechanism and annihilation mechanism, as depicted in Fig.\ref{fig:decay-1} and Fig.\ref{fig:decay-2} respectively. In Table~\ref{tab:decay_mode}, we list some possible two-meson decay modes for these $ud\bar{u}\bar{d}$, $us\bar{u}\bar{s}$ and $ss\bar{s}\bar{s}$ tetraquarks with different $I^G$ quantum numbers. It is clearly that all the final states are P-wave mesons for the S-wave decay modes, while a P-wave plus an S-wave mesons for the P-wave decay modes. For the D-wave decay modes, the final states can be all S-wave mesons, such as the $\rho\pi, \omega\pi, \phi\pi, K\bar K^\ast$ channels. Such peculiar decay properties may result in relative narrow decay widths for these $2^{+-}$ tetraquark states. 

As shown in Fig.\ref{fig:decay-3}, the predicted tetraquark masses in Table~\ref{Tab:Results} also allow some baryon-antibaryon decay channels by the creation of a light quark-antiquark pair, so that the $ud\bar{u}\bar{d}$, $us\bar{u}\bar{s}$ and $ss\bar{s}\bar{s}$ tetraquark states with $J^{PC}=2^{+-}$ may be observed in the $\Delta\bar{\Delta}$, $\Sigma^{\ast} \bar{\Sigma }^{\ast}$, $\Xi^{\ast} \bar{\Xi }^{\ast}$, $\Omega\bar{\Omega }$ decay modes. We suggest to search for these $2^{+-}$ light tetraquark states in the $\rho\pi, \omega\pi, \phi\pi$, $b_{1}\pi$, $h_{1}\pi$, $K\bar K^\ast, K\bar{K}_{1}$, $\Delta\bar{\Delta}$, $\Sigma^{\ast} \bar{\Sigma }^{\ast}$, $\Xi^{\ast} \bar{\Xi }^{\ast}$, $\Omega\bar{\Omega }$ channels in future experiments such as BESIII, BelleII, GlueX, LHCb and so on.

\begin{table} \caption{Some possible two-meson decay modes for the tetraquarks with $I^G(J^{PC})=0^-(2^{+-})$, $1^+(2^{+-})$ and $2^-(2^{+-})$.}
  \renewcommand\arraystretch{1.8} 
  \setlength{\tabcolsep}{1.0em}{ 
\begin{tabular}{cccc}
\hline\hline  
    $I^{G}(J^{PC})$     & $0^{-}(2^{+-})$    &$1^{+}(2^{+-})$ &$2^{-}(2^{+-})$ \\ \hline 
\thead{S-wave}
 & \thead{$K_{0}^{*}\bar{K}_{2}^{*}$,\,\\$a_{1,2}b_{1}$,\,$f_{1,2}h_{1}$}
 & \thead{$K_{0}^{*}\bar{K}_{2}^{*}$,\,\\$a_{1,2}h_{1}$,\,$f_{1,2}b_{1}$} 
 & \thead{$a_{1,2}b_{1}$} \\
\cline{2-4}
\thead{P-wave} 
& \thead{$K\bar{K}_{1}$,\,$K\bar{K}_{2}^{*}$,\,$K^{*}\bar{K}_{0}^{*}$,\,$K^{*}\bar{K}_{1}$,\,$K^{*}\bar{K}_{2}^{*}$,\,\\$h_{1}\eta$,\,$b_{1}\pi$,\,$f_{0,1,2}\omega$\,,\\$f_{0,1,2}\phi$,\,$a_{0,1,2}\rho$} 
& \thead{$K\bar{K}_{1}$,\,$K\bar{K}_{2}^{*}$,\,$K^{*}\bar{K}_{0}^{*}$,\,$K^{*}\bar{K}_{1}$,\,$K^{*}\bar{K}_{2}^{*}$,\,\\ $b_{1}\eta^{(\prime)}$,\,$h_{1}\pi$,\,$f_{0,1,2}\rho$,\,\\$a_{0,1,2}\omega$,\,$a_{1,2}\pi$,\,$b_{1}\rho$}
 & \thead{$b_{1}\pi$,\,$a_{0,1,2}\rho$} \\
\cline{2-4}
\thead{D-wave} 
& \thead{$K\bar{K}^{\ast}$,\,\\$\rho\pi$,\,$\omega\eta^{(\prime)}$,\,$\phi\eta^{(\prime)}$} 
& \thead{$K\bar{K}^{\ast}$,\,\\$\omega\pi$,\,$\rho\eta^{(\prime)}$,\,$\phi\pi$} 
& \thead{$\rho\pi$} \\
\hline\hline  
\end{tabular}}
  \label{tab:decay_mode}
\end{table}

\section*{ACKNOWLEDGMENTS}
This work is supported by the National Natural Science Foundation of China under Grant No.12305147 and No.12175318, the National Key R\&D Program of China under Contracts No. 2020YFA0406400, the Natural Science Foundation of Guangdong Province of China under Grant No. 2022A1515011922.

\appendix
\section*{Appendix: Expressions of correlation functions}\label{appendix}
In this appendix, we show the expressions of correlation functions for the interpolating currents $J_{\alpha \mu \nu }^{1}$, $J_{\alpha \mu \nu }^{1\prime}$, $J_{\alpha \mu \nu }^{2}$, $J_{\alpha \mu \nu }^{2\prime}$ and $J_{\alpha \mu \nu }^{s1}$.
For the nonstrange $ud\bar{u}\bar{d}$ tetraquark system, the correlation functions after Borel transformation are
\begin{equation}
  \begin{aligned}
     \Pi_{d}^{1}(M_{B}^{2},s_{0})=&\int_{0}^{s_{0}} \left(\frac{s^4}{26880 \pi^6}-\frac{\dGG s^2}{1152 \pi^5}-\frac{\dqq^2 s}{6 \pi^2}-\frac{\dqGq \dqq}{9 \pi^2}\right)e^{-\frac{s}{M_{B}^{2}}} ds +  \frac{5 \dqGq^2}{144 \pi^2}+\frac{\dGG \dqq^2}{27 \pi }\, ,
    \end{aligned}
  \end{equation}

\begin{equation}
\begin{aligned}
       \Pi_{d}^{1'}(M_{B}^{2},s_{0})=&\int_{0}^{s_{0}} \left(
       \frac{s^4}{13440 \pi ^6}-\frac{\dGG s^2}{576 \pi ^5}-\frac{\dqq^2 s}{3 \pi ^2}-\frac{5 \dqGq \dqq}{36 \pi ^2} \right)e^{-\frac{s}{M_{B}^{2}}} ds +
      \frac{\dqGq^2}{36 \pi ^2}+\frac{2 \dGG \dqq^2}{27 \pi }\, ,
\end{aligned}
\end{equation}

\begin{equation}
\begin{aligned}
         \Pi_{d}^{2}(M_{B}^{2},s_{0})=&\int_{0}^{s_{0}} \left(
        \frac{s^4}{26880 \pi ^6}-\frac{\dGG s^2}{1152 \pi ^5}-\frac{\dqq^2 s}{6 \pi ^2}-\frac{7 \dqGq \dqq}{72 \pi ^2} \right)e^{-\frac{s}{M_{B}^{2}}} ds +
          \frac{\dqGq^2}{36 \pi ^2}+\frac{\dGG \dqq^2}{27 \pi }\, ,
\end{aligned}
\end{equation}

\begin{equation}
  \begin{aligned}
           \Pi_{d}^{2'}(M_{B}^{2},s_{0})=&\int_{0}^{s_{0}} \left(\frac{s^4}{13440 \pi ^6}-\frac{\dGG s^2}{576 \pi ^5}-\frac{\dqq^2 s}{3 \pi ^2}-\frac{17 \dqGq \dqq}{72 \pi ^2}\right)e^{-\frac{s}{M_{B}^{2}}} ds +
          \frac{11 \dqGq^2}{144 \pi ^2}+\frac{2 \dGG \dqq^2}{27 \pi }\, .
          \end{aligned}
        \end{equation}

For the $us\bar{u}\bar{s}$ tetraquark system, the correlation functions after Borel transformation are
\begin{equation}
\begin{aligned}
   \Pi_{s}^{1}(M_{B}^{2},s_{0})=&\int_{0}^{s_{0}} \left(\frac{s^4}{26880 \pi ^6}+\frac{11 \dss m_{s} s^2}{320 \pi ^4}-\frac{\dGG s^2}{1152 \pi ^5}-\frac{ \dqq^2 s}{12 \pi ^2}-\frac{\dss^2 s}{12 \pi ^2}+\frac{\dqGq m_{s} s}{384 \pi ^4}+\frac{59 \dsGs m_{s} s}{1152 \pi ^4}\right.\\ &\quad \left. -\frac{5 \dqGq \dqq}{96 \pi ^2}-\frac{5 \dsGs \dss}{96 \pi ^2}-\frac{ \dqq \dsGs}{288 \pi ^2}-\frac{\dqGq \dss}{288 \pi ^2}+\frac{7 \dGG \dqq m_{s}}{1152 \pi ^3}\right.\\ &\quad \left. +\frac{\dGG \dss m_{s}}{1152 \pi ^3}\right)e^{-\frac{s}{M_{B}^{2}}} ds
 +  \frac{\dqGq^2}{64 \pi ^2}+\frac{\dsGs \dqGq}{288 \pi ^2}+\frac{\dGG  \dqGq m_{s}}{1152 \pi ^3}\\&\quad-\frac{2\dqq^2 \dss m_{s}}{3} +\frac{\dGG \dqq^2}{54 \pi }+\frac{\dGG \dss^2}{54 \pi }+\frac{\dsGs^2}{64 \pi ^2}+\frac{5 \dGG \dsGs m_{s}}{3456 \pi ^3}\, ,
  \end{aligned}
\end{equation}

\begin{equation}
  \begin{aligned}
     \Pi_{s}^{1'}(M_{B}^{2},s_{0})=&\int_{0}^{s_{0}} \left(\frac{s^4}{13440 \pi ^6}+\frac{11  \dss  m_{s} s^2}{160 \pi ^4}-\frac{\dGG s^2}{576 \pi ^5}-\frac{ \dqq^2 s}{6 \pi ^2}-\frac{ \dss^2 s}{6 \pi ^2}-\frac{7 \dqGq  m_{s} s}{384 \pi ^4}+\frac{109  \dsGs  m_{s} s}{1152 \pi ^4}\right.\\ &\quad \left. -\frac{3 \dqGq  \dqq}{32 \pi ^2}-\frac{3  \dsGs  \dss}{32 \pi ^2}+\frac{7  \dqq  \dsGs}{288 \pi ^2}+\frac{7 \dqGq  \dss}{288 \pi ^2}-\frac{49 \dGG  \dqq  m_{s}}{1152 \pi ^3}\right.\\ &\quad \left. -\frac{19 \dGG  \dss  m_{s}}{1152 \pi ^3}\right)e^{-\frac{s}{M_{B}^{2}}} ds
+\frac{5 \dqGq^2}{192 \pi ^2}-\frac{7  \dsGs \dqGq}{288 \pi ^2}-\frac{7 \dGG\dqGq  m_{s} }{1152 \pi ^3}\\ &\quad  -\frac{4\dqq^2  \dss  m_{s}}{3}  +\frac{\dGG  \dqq^2}{27 \pi }+\frac{\dGG  \dss^2}{27 \pi }+\frac{5  \dsGs^2}{192 \pi ^2}+\frac{\dGG  \dsGs  m_{s}}{3456 \pi ^3}\, ,
    \end{aligned}
  \end{equation}
  \begin{equation}
    \begin{aligned}
    \Pi_{s}^{2}(M_{B}^{2},s_{0})=&\int_{0}^{s_{0}} \left(\frac{s^4}{26880 \pi ^6}+\frac{11 \dss m_{s} s^2}{320 \pi ^4}-\frac{\dGG s^2}{1152 \pi ^5}-\frac{\dqq^2 s}{12 \pi ^2}-\frac{\dss^2 s}{12 \pi ^2}-\frac{\dqGq m_{s} s}{384 \pi ^4}+\frac{59 \dsGs m_{s} s}{1152 \pi ^4}\right.\\ &\quad \left.-\frac{5 \dqGq\dqq}{96 \pi ^2}-\frac{5 \dsGs \dss}{96 \pi ^2}+\frac{\dqq \dsGs}{288 \pi ^2}+\frac{\dqGq \dss}{288 \pi ^2}-\frac{7 \dGG\dqq m_{s}}{1152 \pi ^3}\right.\\ &\quad \left.+\frac{\dGG \dss m_{s}}{1152 \pi ^3}\right)e^{-\frac{s}{M_{B}^{2}}} ds+ \frac{\dqGq^2}{64 \pi ^2}-\frac{\dsGs \dqGq}{288 \pi ^2}-\frac{\dGG\dqGq m_{s} }{1152 \pi ^3}\\ &\quad-\frac{2\dqq^2 \dss m_{s}}{3} +\frac{\dGG \dqq^2}{54 \pi }+\frac{\dGG \dss^2}{54 \pi }+\frac{\dsGs^2}{64 \pi ^2}+\frac{5 \dGG \dsGs m_{s}}{3456 \pi ^3}\, ,
  \end{aligned}
  \end{equation}
  
  \begin{equation}
    \begin{aligned}
    \Pi_{s}^{2'}(M_{B}^{2},s_{0})=&\int_{0}^{s_{0}} \left(\frac{s^4}{13440 \pi ^6}+\frac{11 \dss m_{s}s^2}{160 \pi ^4}-\frac{\dGG s^2}{576 \pi ^5}-\frac{\dqq^2 s}{6 \pi ^2}-\frac{\dss^2 s}{6 \pi ^2}+\frac{7 \dqGq m_{s}s}{384 \pi ^4}+\frac{109 \dsGs m_{s}s}{1152 \pi ^4}\right.\\ &\quad \left.-\frac{3 \dqGq \dqq}{32 \pi ^2}-\frac{3 \dsGs \dss}{32 \pi ^2}-\frac{7 \dqq \dsGs}{288 \pi ^2}-\frac{7 \dqGq \dss}{288 \pi ^2}+\frac{49\dGG \dqq m_{s}}{1152 \pi ^3}\right.\\ &\quad \left.-\frac{19\dGG \dss m_{s}}{1152 \pi ^3}\right)e^{-\frac{s}{M_{B}^{2}}} ds +\frac{5 \dqGq^2}{192 \pi ^2}+\frac{7 \dsGs \dqGq}{288 \pi ^2}+\frac{7\dGG\dqGq m_{s}}{1152 \pi ^3}\\ &\quad -\frac{4 \dqq^2 \dss m_{s}}{3} +\frac{\dGG \dqq^2}{27 \pi }+\frac{\dGG \dss^2}{27 \pi }+\frac{5 \dsGs^2}{192 \pi ^2}+\frac{\dGG \dsGs m_{s}}{3456 \pi ^3}\, .
  \end{aligned}
  \end{equation}
  
For the fully strange $ss\bar{s}\bar{s}$ tetraquark system, the correlation functions after Borel transformation are
    \begin{equation}
    \begin{aligned}
    \Pi_{ss}^{s1}(M_{B}^{2},s_{0})=&\int_{0}^{s_{0}} \left(\frac{s^4}{26880 \pi ^6}+\frac{11 \dss m_{s} s^2}{160 \pi ^4}-\frac{\dGG s^2}{1152 \pi ^5}-\frac{\dss^2 s}{6 \pi ^2}+\frac{31 \dsGs m_{s} s}{288 \pi ^4}-\frac{\dsGs \dss}{9 \pi ^2}\right.\\ &\quad \left.+\frac{\dGG \dss m_{s}}{72 \pi ^3}\right)e^{-\frac{s}{M_{B}^{2}}} ds -\frac{ 4  \dss^3 m_{s}}{3}+\frac{\dGG \dss^2}{27 \pi }+\frac{5 \dsGs^2}{144 \pi ^2}+\frac{\dGG \dsGs m_{s}}{216 \pi ^3}\, .
  \end{aligned}
  \end{equation}

%\bibliography{MyRef}
%\bibliographystyle{plain}
%\bibliographystyle{paper}
%\bibliography{MyRef}

\end{document}